\documentclass[8.5pt,twoside,twocolumn]{article}
\usepackage[utf8]{inputenc}
\usepackage{amsmath,amssymb}
\usepackage[super,sort&compress,comma]{natbib} 
\usepackage{mhchem}
\usepackage{times,mathptmx}
\usepackage{sectsty}
\usepackage{balance} 

\usepackage{graphicx} 
\usepackage{lastpage}
\usepackage[format=plain,justification=raggedright,singlelinecheck=false,font=small,labelfont=bf,labelsep=space]{caption} 
\usepackage{fancyhdr}
\usepackage{color}
\definecolor{darkblue}{rgb}{0,0,0.6}
\definecolor{darkred}{rgb}{0.6,0,0}
\usepackage[colorlinks=true,urlcolor=darkblue,citecolor=darkblue,linkcolor=darkred,hyperfootnotes=false]{hyperref}

\newcommand{\ind}[1]{_{\mathrm{#1}}}

\newcommand{\rr}{\pmb{r}}

\newcommand{\xx}{\pmb{x}}
\newcommand{\zz}{\pmb{z}}

\usepackage{xcolor}

\begin{document}

\thispagestyle{plain}
\fancypagestyle{plain}{
\fancyhead[L]{\includegraphics[height=8pt]{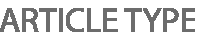}}
\fancyhead[C]{\hspace{-1cm}\includegraphics[height=20pt]{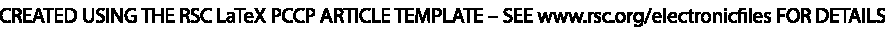}}
\fancyhead[R]{\includegraphics[height=10pt]{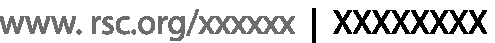}\vspace{-0.2cm}}
\renewcommand{\headrulewidth}{1pt}}
\renewcommand{\thefootnote}{\fnsymbol{footnote}}
\renewcommand\footnoterule{\vspace*{1pt}%
\hrule width 3.4in height 0.4pt \vspace*{5pt}} 
\setcounter{secnumdepth}{5}

\makeatletter 
\def\subsubsection{\@startsection{subsubsection}{3}{10pt}{-1.25ex plus -1ex minus -.1ex}{0ex plus 0ex}{\normalsize\bf}} 
\def\paragraph{\@startsection{paragraph}{4}{10pt}{-1.25ex plus -1ex minus -.1ex}{0ex plus 0ex}{\normalsize\textit}} 
\renewcommand\@biblabel[1]{#1}            
\renewcommand\@makefntext[1]%
{\noindent\makebox[0pt][r]{\@thefnmark\,}#1}
\makeatother 
\renewcommand{\figurename}{\small{Fig.}~}
\sectionfont{\large}
\subsectionfont{\normalsize} 

\fancyfoot{}
\fancyfoot[LO,RE]{\vspace{-7pt}\includegraphics[height=9pt]{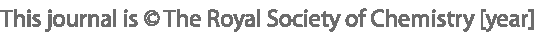}}
\fancyfoot[CO]{\vspace{-7.2pt}\hspace{12.2cm}\includegraphics{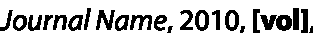}}
\fancyfoot[CE]{\vspace{-7.5pt}\hspace{-13.5cm}\includegraphics{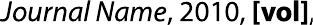}}
\fancyfoot[RO]{\footnotesize{\sffamily{1--\pageref{LastPage} ~\textbar  \hspace{2pt}\thepage}}}
\fancyfoot[LE]{\footnotesize{\sffamily{\thepage~\textbar\hspace{3.45cm} 1--\pageref{LastPage}}}}
\fancyhead{}
\renewcommand{\headrulewidth}{1pt} 
\renewcommand{\footrulewidth}{1pt}
\setlength{\arrayrulewidth}{1pt}
\setlength{\columnsep}{6.5mm}
\setlength\bibsep{1pt}

\twocolumn[
  \begin{@twocolumnfalse}
\noindent\LARGE{\textbf{Mechanics of large folds in thin interfacial films}}
\vspace{0.6cm}

\noindent\large{\textbf{Vincent Démery,$^\ast$ Benny Davidovitch, and Christian D. Santangelo}}\vspace{0.5cm}

\noindent\textit{\small{\textbf{Received Xth XXXXXXXXXX 20XX, Accepted Xth XXXXXXXXX 20XX\newline
First published on the web Xth XXXXXXXXXX 200X}}}

\noindent \textbf{\small{DOI: 10.1039/b000000x}}
\vspace{0.6cm}

\noindent \normalsize{A thin film at a liquid interface responds to uniaxial confinement by wrinkling and then by folding; its shape and energy have been computed exactly before self contact. 
Here, we address the mechanics of large folds, i.e. folds that absorb a length much larger than the wrinkle wavelength. With scaling arguments and numerical simulations, we show that the antisymmetric fold is energetically favorable and can absorb any excess length at zero pressure.
Then, motivated by puzzles arising in the comparison of this simple model to experiments on lipid monolayers and capillary rafts, we discuss how to incorporate film weight, self-adhesion and energy dissipation.
}
\vspace{0.5cm}
 \end{@twocolumnfalse}
]

\footnotetext{\textit{Department of Physics, University of Massachusetts, Amherst, MA 01003, USA.}}
\footnotetext{\textit{$^\ast$~E-mail: vdemery@physics.umass.edu}}



\section{Introduction}\label{}

Deforming soft two dimensional objects by means of capillarity opened a new route to design three dimensional structures at the micro and nanoscale \cite{Py2007,Roman2010,Leong2010}.
However, attaching these thin films to soft substrates submits them to a wealth of morphological instabilities such as wrinkling, crumpling or folding \cite{Li2012}. 
Such instabilities have been observed in lipid monolayers \cite{Ries1979,Milner1989,Saint_Jalmes1998,Ybert2002,Zhang2005,Gopal2006,Pu2006,Lee2008}, nanoparticles films \cite{Leahy2010}, capillary rafts\cite{Vella2004,Protiere2010,Abkarian2013} and thin polymer sheets resting on a gel \cite{Pocivavsek2008,Brau2013} or a liquid substrate \cite{Holmes2010,King2012}.
Their complete characterization is a necessary step towards their control and use in the fabrication of small structures.

\begin{figure}
\begin{center}
\includegraphics[width=\linewidth]{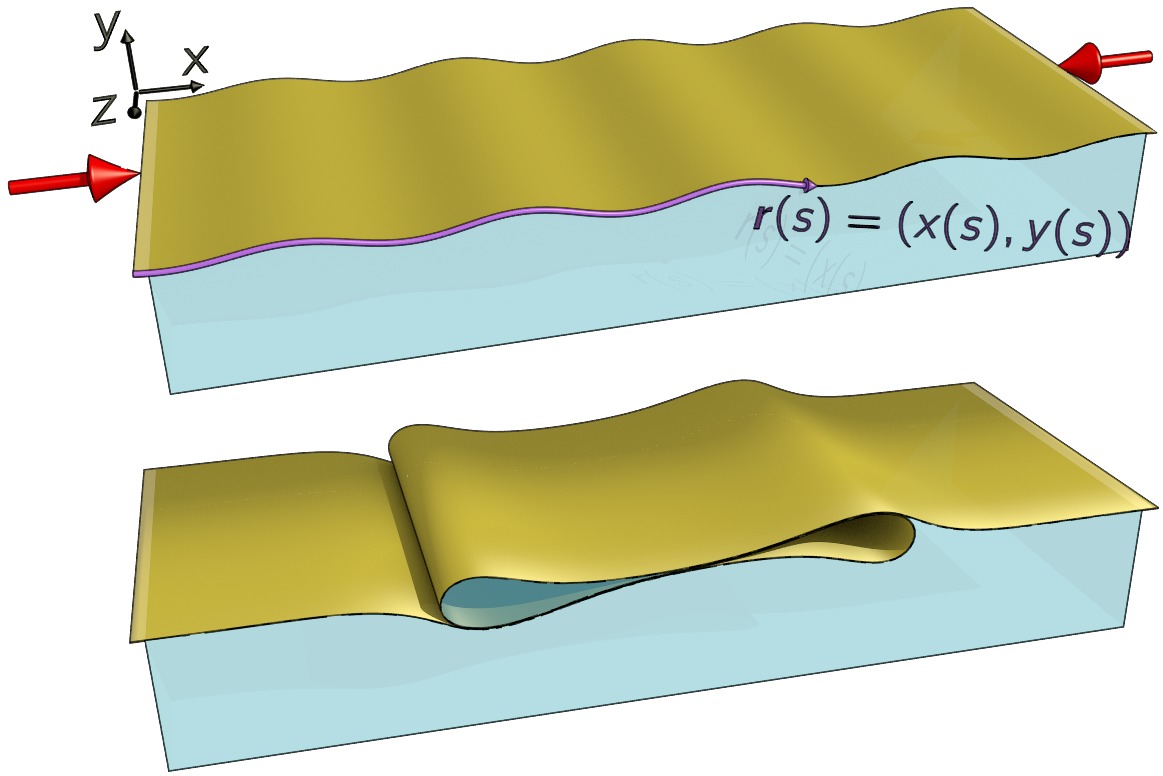}
\end{center}
\caption{A thin interfacial film responds to uniaxial confinement first by wrinkling (top), and then by forming a large fold (bottom). The invariance of the system along $\hat\zz$ allows to parametrize the shape of the film by a function $\rr(s)=(x(s),y(s))$.}
\label{fig:picture}
\end{figure}

A simple setup where some of these instabilities arise consists of a thin film at an initially flat liquid interface that is confined in one horizontal direction (see Fig.~\ref{fig:picture}).
The film responds to confinement by wrinkling and folding in a universal way resulting from the competition between the bending energy to deform the film  and the gravitational energy to lift the liquid\cite{Pocivavsek2008}. Minimizing these energies leads to an integrable equation for the shape of the film\cite{Diamant2011,Rivetti2013,Diamant2013}, allowing one to obtain an analytical expression for the energy as a function of the confinement length. However, this exact result holds only up to self-contact of the film, and that occurs as soon as the confinement length reaches approximately the wavelength of the wrinkles.

On the other hand, the experimental range of confinement for lipid monolayers\cite{Ries1979,Milner1989,Zhang2005,Gopal2006,Pu2006,Lee2008} and capillary rafts\cite{Protiere2010,Abkarian2013} goes far beyond self contact.
In the first case, folds are formed\cite{Ries1979} abruptly, causing jerky monolayer dynamics\cite{Gopal2006}. In a folding event, a length $\sim 2\,\mu\textrm{m}$ is absorbed in a fold in $\sim 0.1\,\textrm{s}$. 
It has already been noted that this characteristic time is anomalously fast\cite{Oppenheimer2013}, but what sets the characteristic length is also unclear.
In capillary rafts, large folds -- involving a length much larger than the wrinkle wavelength -- are formed and eventually get destabilized under their own weight\cite{Protiere2010}. 
To understand the behavior of the film in these experiments, a systematic study of the mechanics of large folds is required. 
In this article, we address the following questions: what is the shape of a fold after self contact? What is its energy?

\section{Model}\label{}

A thin film at a liquid interface is submitted to uniaxial confinement along $\hat\xx$; the system is invariant in the $\hat\zz$ direction (see Fig.~\ref{fig:picture}). 
Soon after the confinement length exceeds a threshold value for wrinkling instability, the film responds as if it was nearly inextensible, and can be modeled as a rod parametrized by $\rr(s)=(x(s),y(s))$ in a vertical plane ($s$ is the arc length).
\citet{Pocivavsek2008} found that the bending energy of the film and the gravitational energy of the displaced fluid are responsible for the wrinkle to fold transition. Those energies are, respectively,
\begin{align}
U\ind{bend} & = \frac{B}{2}\int \rr''(s)^2ds, \label{eq:ubend}\\
U\ind{grav} & = \frac{\rho g}{2}\int y(s)^2x'(s)ds, \label{eq:ugrav}
\end{align}
where $B$ is the bending modulus of the film,
 $\rho$ is the mass density difference between the fluids below and above the sheet, $g$ is the gravitational acceleration and energies are given per unit length in the orthogonal direction.
For a continuous material, the bending modulus is given by $B=Et^3/[12(1-\nu^2)]$, where $E$ is the Young modulus, $\nu$ the Poisson ratio, and $t$ the thickness of the film.
These parameters allow one to define the characteristic length $l=(B/\rho g)^{1/4}$. 
Rescaling lengths by $l$ and energies by $B/l$, we are left with a system with no dimensionless parameters (in the following, we use only dimensionless quantities). 
We focus on the dependence of the energy on the confinement length $\Delta=L-[x(L)-x(0)]$, $L$ being the length of the film in the confined direction.

\begin{figure}
\begin{center}
\includegraphics[width=\linewidth]{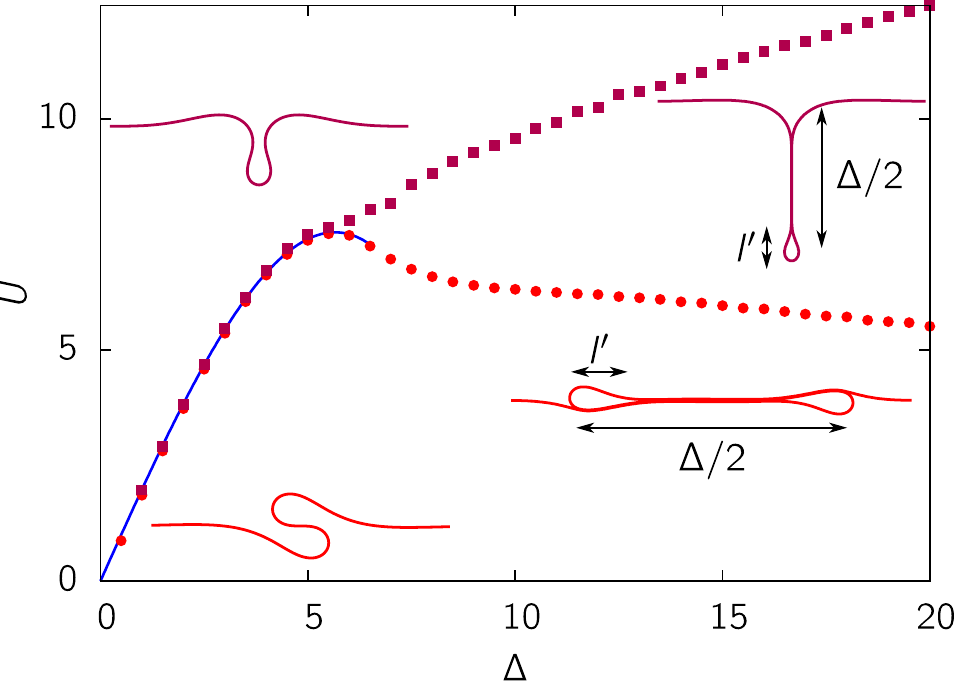}
\end{center}
\caption{Fold energy as a function of the imposed displacement for the symmetric (squares) and antisymmetric (circle) folds. The blue line is the exact solution (\ref{eq:exact}) valid before self-contact.
Symmetric and antisymmetric configurations are shown before self contact (left, exact solutions from \citet{Diamant2011}) and after self contact (right). After self contact, the size of the fold $\Delta/2$ absorbs the excess length while bending is localized in highly curved zones of length $l'$.}
\label{fig:d_u_sym_asym}
\end{figure}

The system defined by the energies (\ref{eq:ubend}-\ref{eq:ugrav}) is integrable \cite{Diamant2011,Rivetti2013,Diamant2013}. For a given confinement length $\Delta$, there is a continuous family of solutions with the same energy 
\begin{equation}\label{eq:exact}
U^0(\Delta)=2\Delta-\frac{\Delta^3}{48},
\end{equation}
among which are the symmetric and antisymmetric configurations pictured in Fig.~\ref{fig:d_u_sym_asym}.
Two points are noteworthy: first, this energy is always lower than the energy of the wrinkled state, $U^\mathrm{wr}=2\Delta$~\cite{Milner1989}; second, this energy has a maximum and may even become negative. This is prevented by self contact, where the exact solutions cease to be valid. Self contact occurs at $\Delta\simeq 5.6$ for the symmetric fold, just before the maximum, and at $\Delta\simeq 6.6$ for the antisymmetric fold, just after $U^0$ reaches its maximum, meaning that there exists an antisymmetric fold with negative pressure.

\section{Scaling arguments and numerical solution}\label{}

We start our investigation of large folds with a scaling analysis.
Large symmetric and antisymmetric folds are depicted in Fig.~\ref{fig:d_u_sym_asym}. 
A fold is characterized by two lengths: its size $\Delta/2$ that is given by the confinement length (assuming that the whole confinement length is absorbed into the fold), and the size $l'$ of the highly curved zone(s) that contains bending. The size $l'$ is determined by an energy balance.
In the symmetric case, the bending and gravitational energies are respectively $1/l'$ and $\Delta l'^2$ (the volume of fluid contained in the highly curved zone is $l'^2$ and its displacement is $\Delta$). Minimizing over $l'$ gives $l'\sim \Delta ^{-1/3}$ and the scaling law
\begin{equation}\label{eq:asymptotic_sym}
U^\mathrm{sym}\sim\Delta^{1/3}.
\end{equation}
Note that this scaling is strictly different from the result of a scaling argument in \citet{Pocivavsek2008}, which neglected the effect of self avoidance. 
On the other hand, in the antisymmetric case, the displacement of the fluid inside the highly curved zones does not depend on the fold size $\Delta$. The bending and gravitational energies are, respectively, $1/l'$ and $l'^3$, leading to $l'\sim 1$ and 
\begin{equation}\label{eq:asymptotic_asym}
U^\mathrm{antisym}\sim 1.
\end{equation}
Since the fold occurs at $\Delta>1$, the antisymmetric fold has a lower energy than the symmetric one, which does not depend on the size of the fold: once it is formed, it can absorb length at negligible cost.

In order to completely characterize the behavior of the fold, we have to investigate the crossover between the energy at self contact, given by Eq.~\ref{eq:exact}, and the asymptotic behaviors of Eqs.~\ref{eq:asymptotic_sym}-\ref{eq:asymptotic_asym}. Besides this crossover, we want to determine the asymptotic value of the energy of the asymmetric fold. 
We resort to a numerical computation of the film shape to answer these questions.

The rod parametrized by $\rr(s)$ is modeled as a chain of beads
with bending and gravitational energies, a stretching energy with a very large stretching modulus and a short-range repulsion energy between the beads to prevent self-crossing.
The equilibrium rod configuration is given by minimization of the full energy, and its energy is computed with the bending and gravitational contributions only.
We perform two kinds of simulations: in the first, we find the energy minimizing configuration of the complete rod; in the second, we consider one half of the rod and impose a symmetric configuration.
In the first case, the energy minimizing configuration is always found to be antisymmetric after self-contact.

The energies of the symmetric and antisymmetric configurations are plotted as a function of the displacement in Fig.~\ref{fig:d_u_sym_asym}: they are equal before self-contact and differ strongly after self-contact. The energy of the symmetric fold keeps increasing while that of the antisymmetric fold decreases monotonously to a plateau well below its maximum value (the symmetric fold shown in Fig.~\ref{fig:d_u_sym_asym} is pointing down, but the corresponding configuration with the fold pointing up has the same energy).

\begin{figure}
\begin{center}
\includegraphics[width=\linewidth]{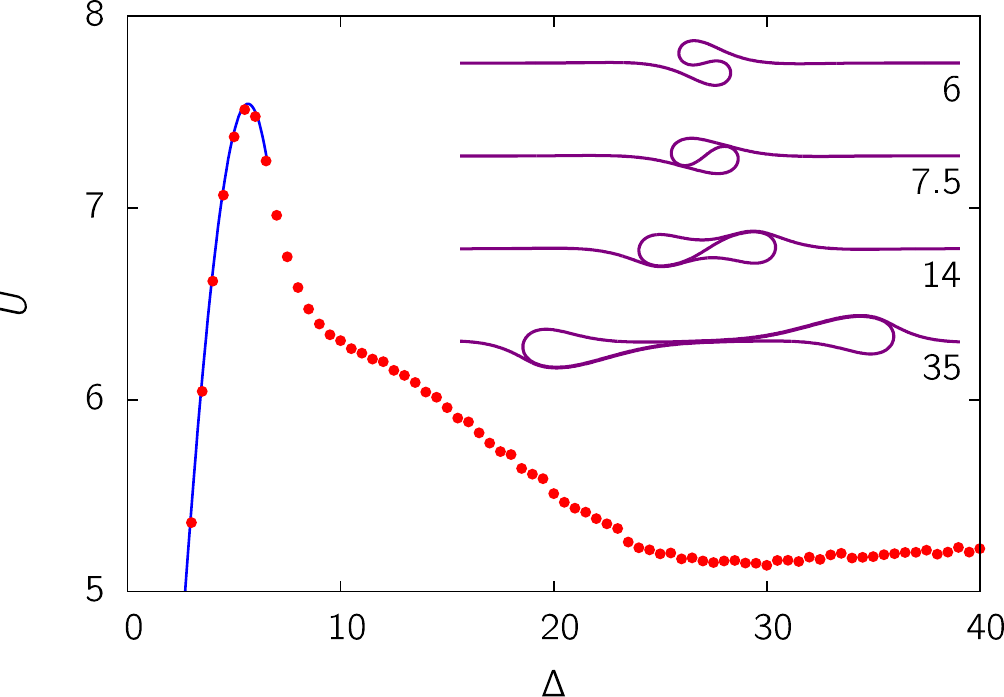}
\end{center}
\caption{Fold energy as a function of the imposed displacement. The blue line is the exact solution (\ref{eq:exact}) that is valid before self contact, i.e. for $\Delta\lesssim 6.6$.
Inset: typical configurations showing the different folding steps (numbers indicate the imposed displacement).}
\label{fig:num_asym}
\end{figure}

The energy of the antisymmetric fold reaches its maximum $U\ind{max}=16\sqrt{2}/3\simeq 7.5$ before self-contact and then decreases to its plateau value $U_\infty\simeq 5.2 \simeq 0.7U\ind{max}$ (see~Fig.~\ref{fig:num_asym}). 
Four steps can be identified after the energy maximum, pictured in the inset of Fig.~\ref{fig:num_asym}: (i) Between the maximum and self-contact ($4\sqrt{2}\simeq 5.7\leq\Delta\leq 6.5$), the two highly curved zones get closer, reducing the gravitational energy.
(ii) Just after self-contact ($6.5\leq\Delta\leq 9$) the size of the highly curved zones increases.
(iii) Once the highly curved zones have reached their optimal size, they start to move apart until a trilayer is formed between them ($9\leq\Delta\leq 25$).
(iv) The highly curved zones move apart at constant energy and shape, increasing the trilayer length between them ($\Delta\geq 25$).

The transition from the flat to the folded film resembles the monolayer to trilayer transition observed in nanospheres rafts\cite{Leahy2010}.

\section{Discussion}\label{}

The picture that emerges from our analysis is that large folds are antisymmetric and energetically cheap; it does not fit the observations of folds in capillary rafts or in lipid monolayers:
\begin{itemize}
\item folds in capillary rafts are often symmetric\cite{Protiere2010,Abkarian2013};
\item folds in lipid monolayers have a well defined length, i.e., creating several folds is favorable to enlarging one fold\cite{Gopal2006}.
\end{itemize}
These discrepancies indicate that the energetic consideration of bending and gravity alone do not account for the observed properties: other effects should be involved.
%
As a preliminary exploration of such effects, we discuss possible extensions of the simple model used here and their potential effect on the energy and dynamics of large folds. 

The film weight plays a crucial role in capillary rafts, leading to fold instability and breaking\cite{Protiere2010,Abkarian2013}. Here, we discuss its effect on the shape and energy of the fold.
The film weight enters in the energy via an additional term
\begin{equation}
U\ind{weight}=M \int y(s)\,ds,
\end{equation}
where $M=\rho\ind{f}[g/(B\rho^3)]^{1/4}$ and $\rho\ind{f}$ is the effective mass of the film per unit area (taking into account its buyoancy). 
It is noteworthy that for the very small deformations involved in the wrinkled phase, the gravitational energy is approximated by $U\ind{grav}\simeq(1/2)\int y(s)^2\,ds$, thus the film weight can be absorbed in a shift of the $y$ coordinate and has no effect.
For large folds, it is straightforward to incorporate it into the scaling analysis: it contributes to the downward symmetric fold (that is selected if $\sigma>0$) as $U\ind{weight}\sim -M \Delta^2$ and it does not contribute to the antisymmetric fold.
This negative contribution may thus make the symmetric fold favorable and even unstable since its energy $U^\mathrm{sym}\sim \Delta^{1/3}-M\Delta^2$ decreases to $-\infty$ after its maximum at $\Delta\ind{c}\sim M^{-3/5}$.
Moreover, a tension $T\sim\Delta$ is induced in the film that will eventually break; 
this behavior is observed in heavy capillary rafts~\cite{Protiere2010,Abkarian2013}.
On the other hand, for monolayers, a rough estimate gives $M\simeq 10^{-3}$ in dimensionless units, meaning that the weight of the film may have an effect only when $\Delta\ind{c}\simeq 100$, i.e. for very large folds. A strong effect of the weigth of the monolayers on their folding is thus unlikely.

We turn to self-attraction, that can hold two parts of the fold together\cite{Holmes2010} and has been suggested as a mechanism to drive the folding of lipid monolayers\cite{Lee2008}.
It can be modelled by an energy gain $\Gamma$ per unit area of the film in contact with itself.
The adhesion energy $\Gamma$ may differ on either side of the film: not only the two sides of the film can be different, as is the case for lipid monolayers, but the interaction of the film with itself can depend on the surrounding liquid.
The symmetric fold is the first to experience self-contact, thus it may be favored in the presence of self attraction. Self attraction leads to an energy gain $\Gamma\Delta$, giving $U^\mathrm{sym}\sim\Delta^{1/3}-\Gamma\Delta$. Thus, depending on $\Gamma$, the film may be unstable at self contact. 
In case self-adhesion prevents relative motion and fluid flow between two sections of the film in contact with each other, the size of the highly curved zone cannot decrease as $l'\sim\Delta^{-1/3}$ (it requires fluid flow from the highly curved zone to the upper reservoir) and remains equal to its value at self contact, $l'\sim 1$, resulting in the total energy $U^\mathrm{sym}\sim(1-\Gamma)\Delta$.
Let us now consider the antisymmetric fold, with self-attraction on the two sides: the energy gain is higher than in the symmetric case (although self contact occurs later), but relative motion of sticking parts is required, thus, if the upper fluid is sufficiently viscous, the symmetric fold is preferable.
If only one side experiences self attraction, the relative motion of sticking parts can be avoided, the energy gain is the same as in the symmetric case but the gravitational cost (of the liquid phase) is lower: the antisymmetric fold is still favored. 

Lastly, energy dissipation may occur during the fold formation due to flow between nearly touching parts of the film (symmetric fold) or the relative motion of nearly touching parts of the film (antisymmetric fold).
When the symmetric fold grows, the highly curved zone shrinks as $l'\sim\Delta^{-1/3}$ under the effect of increasing hydrostatic pressure and an upward flow is generated in the narrow neck formed by the two parts of the film that are close to self contact. The radius of the highly curved zone shrinks slowly, thus the dissipation decreases as the fold size increases.
In the antisymmetric fold, the effect is slightly different: the length of the highly curved zones does not change, but proximal parts of the film are in relative motion (in Fig.~\ref{fig:num_asym}, inset $\Delta=35$, the top part of the trilayer moves left, the center part does not move and the bottom part moves right). A flow is needed to lubricate this relative motion, and the dissipation increases as $\Delta$, the length of the portions of the film in self contact.
Hence, energy dissipation will be lower in the symmetric fold, and hence it is favored if the formation of the fold is rapid. Once formed, the symmetric fold will eventually relax to the antisymmetric configuration.
A more precise analysis of the sources of energy dissipation would consider the effect of self-attraction, that may reduce the thickness of the fluid layer between parts of the film and thus increase dissipation.

\section{Conclusion}\label{}

We have investigated the behavior of large folds that may appear in thin interfacial films under uniaxial confinement. 
Under the assumption that the system is controlled by bending and gravity~\cite{Pocivavsek2008}, we have shown that the large folds are antisymmetric and their energy decreases after a maximum reached before self contact to a universal value well below this maximum (see Fig.~\ref{fig:num_asym}). The antisymmetric folds are energetically cheap -- one fold can absorb all the excess length at a finite cost -- and  stable -- they do not unfold spontaneously at zero tension.
On the other hand, the energy of symmetric folds increases monotonously, and these folds are thus less favorable energetically.

Although antsymmetric folds may be the actual cause of ``tri-layers", which have been observed, for instance, by \citet{Leahy2010}, their actual development for the wrinkled state had not been directly observed.
We have shown that they do not explain
 the preferred fold size observed in compressed lipid monolayers\cite{Gopal2006}. Together with the kinetic puzzle encountered in trying to predict the folding timescale of monolayers~\cite{Oppenheimer2013}, this suggests that other interactions must be included in the model. 
We discussed the effect of the weight of the film, its self-attraction, and energy dissipation,
finding that the symmetric fold may be favored in some cases. A more thorough study of these effects is however needed to draw quantitative predictions on the modifications of the folding behavior presented here.

On the other hand, Rivetti and Antkowiak\cite{Rivetti2013b,Rivetti2013} have observed the exact solutions of the model presented here~\cite{Diamant2011,Rivetti2013,Diamant2013}. 
The large size system -- the characteristic length is $l\sim 1\,\text{cm}$ -- used in their experiment appears to be accurately described by bending and gravity only, and is thus likely to exhibit the folding behavior predicted here.

\section*{Acknowledgements}
V.D. thanks S. Proti\`ere and M. Abkarian for stimulating discussions about the shape and stability of folds appearing in confined granular rafts, and A. R. C. Romaguera and J. Paulsen for useful advices on the numerical computations. 
The authors acknowledge financial support by the KECK foundation Award 37086 (V.D.), and NSF CAREER Award DMR-11-51780 (B.D.).




\footnotesize{
\bibliography{biblio} 
\bibliographystyle{rsc} 
}

\end{document}